\documentclass[conference]{IEEEtran}

\def\BibTeX{{\rm B\kern-.05em{\sc i\kern-.025em b}\kern-.08em
    T\kern-.1667em\lower.7ex\hbox{E}\kern-.125emX}}

\usepackage{graphicx}
\usepackage{bm}
\usepackage{subcaption}
\usepackage{threeparttable}
\usepackage{amsmath,amssymb,amsfonts}
\usepackage{multicol}
\usepackage{pifont}
\usepackage{xcolor}
\newcommand{\norm}[1]{\left\lVert#1\right\rVert}
\usepackage{booktabs}
\usepackage[Symbol]{upgreek}
\usepackage{refcount}
\ifCLASSOPTIONcompsoc
  \usepackage[nocompress]{cite}
\else
  \usepackage{cite}
\fi
\ifCLASSINFOpdf
\else
\fi
\begin{document}
%
\title{Attentive Autoencoders for Multifaceted Preference Learning in One-class Collaborative Filtering}


\author{\IEEEauthorblockN{Zheda Mai\textsuperscript{\textsection},
Ga Wu\textsuperscript{\textsection}, Kai Luo and
Scott Sanner}
\IEEEauthorblockA{Department of Mechanical \& Industrial Engineering, University of Toronto, Canada\\
zheda.mai@mail.utoronto.ca, 
\{wuga, kluo, ssanner\}@mie.utoronto.ca,
}}

%


\maketitle

\begingroup\renewcommand\thefootnote{\textsection}
\footnotetext{Equal contribution}
\endgroup
\begin{abstract}
Most existing One-Class Collaborative Filtering (OC-CF) algorithms estimate a user's preference as a latent vector by encoding their historical interactions. However, users often show diverse interests, 
which significantly increases the learning difficulty. In order to capture multifaceted user preferences, existing recommender systems either increase the encoding complexity or extend the latent representation dimension. Unfortunately, these changes inevitably lead to increased training difficulty and exacerbate scalability issues. In this paper, we propose a novel and efficient CF framework called Attentive Multi-modal AutoRec (AMA) that explicitly tracks multiple facets of user preferences. Specifically, we extend the Autoencoding-based recommender AutoRec to learn user preferences with multi-modal latent representations, where each mode captures one facet of a user's preferences. By leveraging the attention mechanism, each observed interaction can have different contributions to the preference facets. Through extensive experiments on three real-world datasets, we show that AMA is competitive with state-of-the-art models under the OC-CF setting. Also, we demonstrate how the proposed model improves interpretability by providing explanations using the attention mechanism.
\end{abstract}


%
\IEEEpeerreviewmaketitle

\section{Introduction}

Collaborative Filtering (CF) is one of the most prevalent approaches for personalized recommendations that only relies on historical user-item interaction data. 
Many CF systems in the literature perform well when provided with explicit user feedback such as 1-5 user ratings of a product. In many applications, however, only positive feedback is available, such as view counts of a video or purchases of an item. 
One challenge of working with implicit feedback data is the lack of negative signals. Although we can safely presume that the purchases indicate the preference of a customer to a product, we should not construe the unobserved interactions as negative feedback since a customer would not be aware of all the items on a platform and could not purchase every single item they like. The recommendation setting with only positive feedback observed is known as One-Class Collaborative Filtering~(OC-CF)~\cite{pan2008one}.

In the OC-CF setting, many CF algorithms learn the preference representation of each user as a latent vector by encoding their historical interactions. However, users often have diverse preferences,
which significantly increases the difficulty of modeling multifaceted preferences. In order to capture complicated user preferences, existing recommender systems can either increase the encoding complexity to enhance the modeling capability or extend the latent representation to a higher dimension. Unfortunately, models with complex encoding modules are fraught with increased training difficulty, while models with high dimensional latent spaces require strong regularization to prevent overfitting and prove challenging to scale to large, real-world datasets. 

\begin{figure}
  \centering
  \includegraphics[width=0.4\textwidth]{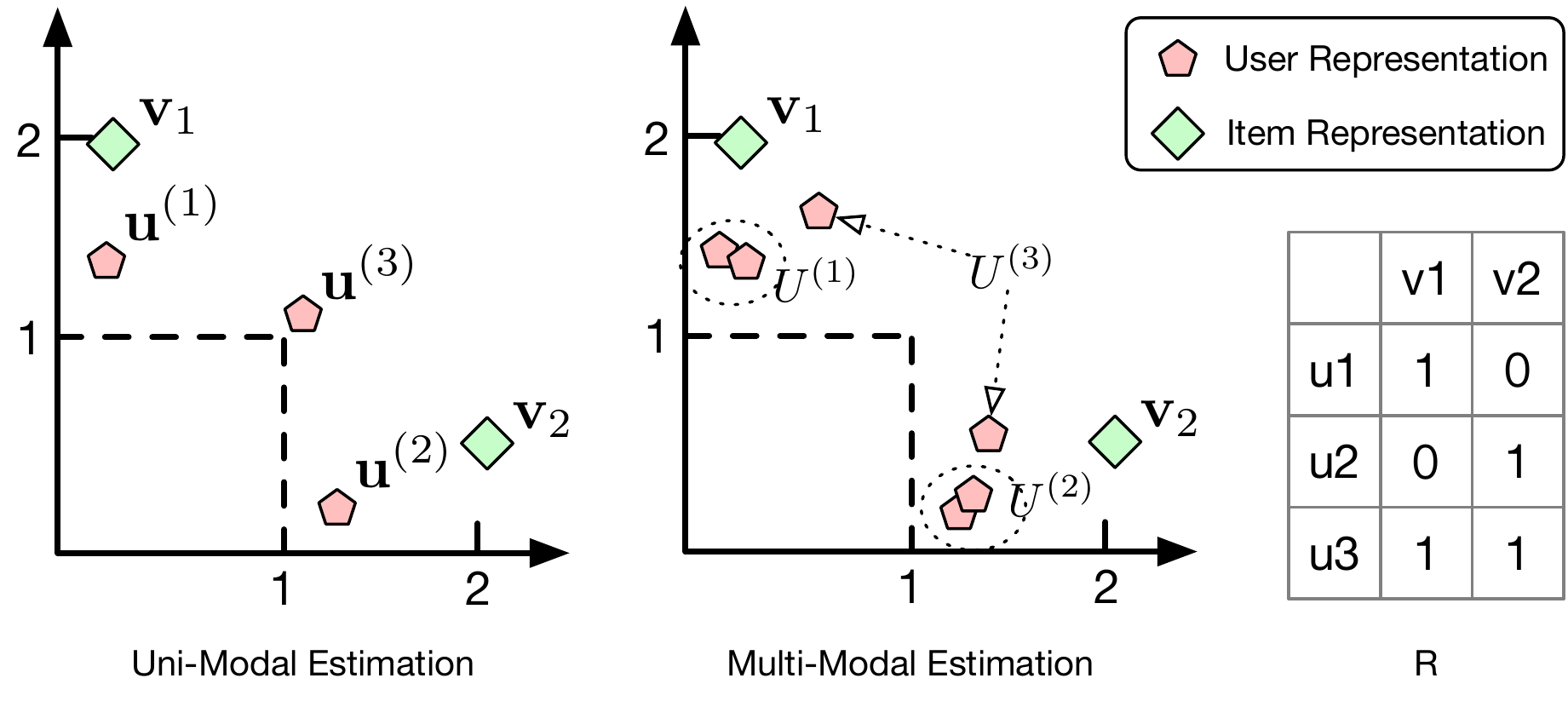}
  \caption{A 2-dimensional latent space conceptualization of uni-modal user preference estimation vs. multi-modal user preference estimation. Uni-modal user-preference estimation results in poor performance because of the difficulty in capturing common properties of user-item interactions. Multi-modal user-preference estimation mitigates the issue by learning several significant potential preference modes.}
  \label{fig:mmp}
\end{figure}

However, learning complex user preferences in low-dimensional latent spaces is not easy.  Most popular recommender systems aim to model user preferences as either a single point or a uni-modal distribution in the low-dimensional latent representation space, which we denote as {\it uni-modal estimation}. This approach can capture the preference for users who have a single preference type. However, for users with multifaceted preferences, the uni-modal estimation may not suffice since it is hard to capture the characteristics of all user preferences with one mode. As we can see in Figure~\ref{fig:mmp}, for users with multifaceted preferences, the uni-modal estimation will try to "average" the user-item interactions and lead to an inaccurate estimation (Figure~\ref{fig:mmp}). Therefore, we should model user preferences with more than one mode in the latent space, which we denote as { \it multi-modal estimation}~{\footnote{The term multi-modal in our context means the latent representation has multiple modes in the latent space. Specifically, it is different from the one in multi-modal learning~\cite{NIPSmultimodal}, where data consists of multiple input modalities.}}. With this approach, each mode in the latent representation explicitly captures one facet of the user preference and can better handle the multifaceted preference situation we mentioned (Figure~\ref{fig:mmp}).

Nevertheless, directly modeling multifaceted preferences is still non-trivial as the ground-truth mappings between item interactions and the facets of user preferences are not available, and individual item interactions may contribute differently to different facets.  To address this problem, we propose a novel and efficient CF framework called Attentive Multi-modal AutoRec (AMA) to track multifaceted user preferences explicitly. AMA is based on AutoRec \cite{sedhain2015autorec}, one of the earliest Autoencoder based recommender systems that aims to learn a low dimensional latent representation that can reconstruct users' observed interactions. But traditional AutoRec adopts a {\it uni-modal estimation} approach and uses a single embedding vector to represent diverse user preferences.

To capture multifaceted user preferences, we replace the uni-modal estimation in AutoRec with multi-modal estimation, where we have more than one latent embedding for each user (i.e., one for each preference facet). Additionally, we take advantage of the attention mechanism to assign different weights to each item interaction when we estimate different latent embeddings. During the decoding phase, each latent embedding will create a prediction, and the model will select the prediction with the highest output value. Apart from the benefit of capturing multifaceted user preferences, AMA can also provide a reasonable interpretation for the recommendations since we know which latent embedding a recommendation comes from and which items contribute to the corresponding latent embedding.  We evaluate AMA extensively on three real-world datasets, and the results show AMA is competitive with many state-of-the-art models with much fewer parameters.

Our contributions are summarized as follows:
\begin{itemize}
\item We propose a novel and efficient Autoencoder-based recommender method called Attentive Multi-modal AutoRec (AMA) to accommodate users with multifaceted preferences in OC-CF. 

\item We employ the attention mechanism to generate multi-modal latent embeddings to capture multifaceted user preferences as well as provide a reasonable interpretation of the recommendations with a localized k-nearest neighbor method.

\item Through extensive experiments on three real-world datasets, we demonstrate that AMA is competitive with many state-of-the-art methods, albeit with many fewer parameters and much lower computational requirements.
\end{itemize}

\section{Preliminary}
\subsection{Notation}
Before proceeding, we define our notation as follows:
\begin{itemize}
    \item $\mathbf{r}^{(i)}$: The implicit feedback vector of user $i$ in shape of $n \times 1$ where $n$ is the number of items. Each entry of $r_{j}^{(i)}$ is either 1, which indicates there is an interaction between user $i$ and item $j$, or 0 otherwise (no interaction).
    \item $\hat{\mathbf{r}}^{(i)}$: Reconstructions of the original vector $\mathbf{r}^{(i)}$.
    \item $\tilde{\mathbf{r}}^{(i)}$: Randomly corrupted implicit feedback vector $\mathbf{r}^{(i)}$ whose entries are randomly reset to 0.
    \item $\mathbf{v}_j$: Latent embedding vector of item $j$ in shape of $h \times 1$ where $h$ is the size of embedding.
    \item $\tilde{\mathbf{v}}_j$: Value vector of item $j$ in attention with shape of $h \times 1$.
    \item $U^{(i)}$: Latent embedding matrix of user $i$ in shape of $d \times h$ where $d$ is the number of user embedding vector. $\mathbf{u}_l^{(i)}$ denotes the $l^{th}$ latent embedding vector for user $i$ with shape of $h \times 1$.
    
    \item $\mathbf{c}^{(i)}$: Loss weighting vector of user $i$. We will define in the paper shortly.
\end{itemize}

\subsection{AutoRec}
AutoRec~\cite{sedhain2015autorec} is a state-of-the-art Autoencoder-based collaborative filtering system. It
embeds a user's sparse preference observations in a latent space and  reconstructs a dense version of those preferences from the embedding to enable  personalized recommendations.

The architecture of the original AutoRec is straightforward -- a fully connected neural network with one hidden layer and sigmoid activation function, which has no difference with conventional Autoencoders. Formally, 
\begin{equation}
    \mathbf{u}^{(i)} = f_\vartheta(\mathbf{r}^{(i)}) \quad \textit{and} \quad \hat{\mathbf{r}}^{(i)} = f_\theta(\mathbf{u}^{(i)}).
\end{equation}
where $\mathbf{u}^{(i)}$ is the user representation vector that encodes user preferences, and $\vartheta$, $\theta$ are the weights of encoder and decoder networks, respectively. 

However, since unobserved interactions do not deliver information about user preferences, AutoRec modifies its objective function as follows:
\begin{equation}
    {\arg\min}_{\theta,\vartheta}\sum_i^m\norm{\mathbf{r}^{(i)} - f_{\theta}(f_{\vartheta}(\mathbf{r}^{(i)}))}_{\mathcal{O}}^2 + \frac{\lambda}{2}(\norm{\theta}_F^2 + \norm{\vartheta}_F^2),
\end{equation}
where $\norm{\cdot}_{\mathcal{O}}^2$ means that the training only takes observed ratings into consideration.

The performance of Autoencoder-based recommender systems highly relies on the accuracy of the user preference embedding, where the preferences of a user could be diverse. In order to better capture complicated user preferences, many previous works propose to increase model complexity through either expanding the latent representation space or training deeper encoder/decoder neural networks (see \cite{kuchaiev2017training,cao2017online,zhuang2017representation}). Unfortunately, these models must trade off between model complexity and data sparsity in recommendation tasks. Specifically, complex models overfit the training data due to insufficient observed user-item interactions. In contrast, simple models fail to capture fine-grained user preferences and introduce biases in prediction (e.g., popularity bias).

\subsection{Attention Mechanism}
\label{sec:attention_mechanism}
The attention mechanism in a deep neural network is a knowledge retrieval method that automatically 
computes a latent representation $\mathbf{z}$ from various weighted source features $V = [\mathbf{v}_1\cdots\mathbf{v}_n]$ with attention weights $\mathbf{a}$ as follows:
\begin{equation}
    \mathbf{z} = \sum_{j=1}^{n} a_j\mathbf{v}_j.
\end{equation}
The distribution of the attention weights $\mathbf{a}$ reflects quantified interdependence 
between a query $\mathbf{q}$ and a list of 
feature keys $K = [\mathbf{k}_1\cdots\mathbf{k}_n]$. Usually, higher attention weight $a_j$ indicates the higher value of the corresponding feature $\mathbf{v}_j$ to the task. 

The most commonly used attention mechanism is Scaled Dot-Product Attention introduced along with the Transformer architecture \cite{vaswani2017attention}, as defined in Equation~\ref{eq:attention_mechanism}:

\begin{equation}
\mathbf{a} = \operatorname{softmax}\left(\frac{ K\mathbf{q}}{\sqrt{\kappa}}\right),
\label{eq:attention_mechanism}
\end{equation}
where $\kappa$ denotes the length of key and query vectors.

\subsection{Attentive Recommender Systems}
Many recent research shows the attention mechanism could significantly improve recommendation performance~(see \cite{chen2017attentive,li2017neural,zhou2018atrank,kang2018self,fu2018attention}). Especially, in content-based recommender systems, the attention mechanism aims to extract features from various content sources and is proven to be effective. Attentive Collaborative Filtering~(ACF)~\cite{chen2017attentive} is one of the most expressive model representations, which extracts informative components from 
the multimedia items.

In addition, the attention mechanism also contributes to session-based recommendation tasks, where the recommender system actively tracks a users' preferences as they dynamically evolve. 
ATRank~\cite{zhou2018atrank} is a representative model for session-based recommenders, which adopts self-attention to model heterogeneous observations of user behaviour.

Unfortunately, less effort has been made on incorporating the attention mechanism into traditional collaborative filtering tasks, where the system is only provided with the 
user-item interaction history.
Latent Relational Metric Learning~(LRML)~\cite{tay2018latent} is a pioneering collaborative filtering approach that uses attention to model the relationship between user-item interactions. 
However, similar to the other Metric Learning based CF algorithms, LRML suffers from exceptionally slow convergence due to negative sampling.

We note that some content-based filtering systems could support collaborative filtering tasks with negligible changes by replacing content-based item-embeddings with collaborative item-embeddings. For example, we can obtain the collaborative item-embeddings simply from SVD decomposition of the interaction matrix $R$. Through this method, content-based systems such as Attentive Collaborative Filtering~(ACF)~\cite{chen2017attentive} are capable of tackling  collaborative filtering tasks without side information. However, as those systems are not intentionally invented for pure collaborative filtering tasks, the advantage of their sophisticated architecture design turns out to be a burden when encountering highly sparse observations.
\begin{figure}
    \centering
    \includegraphics[width=0.3\textwidth]{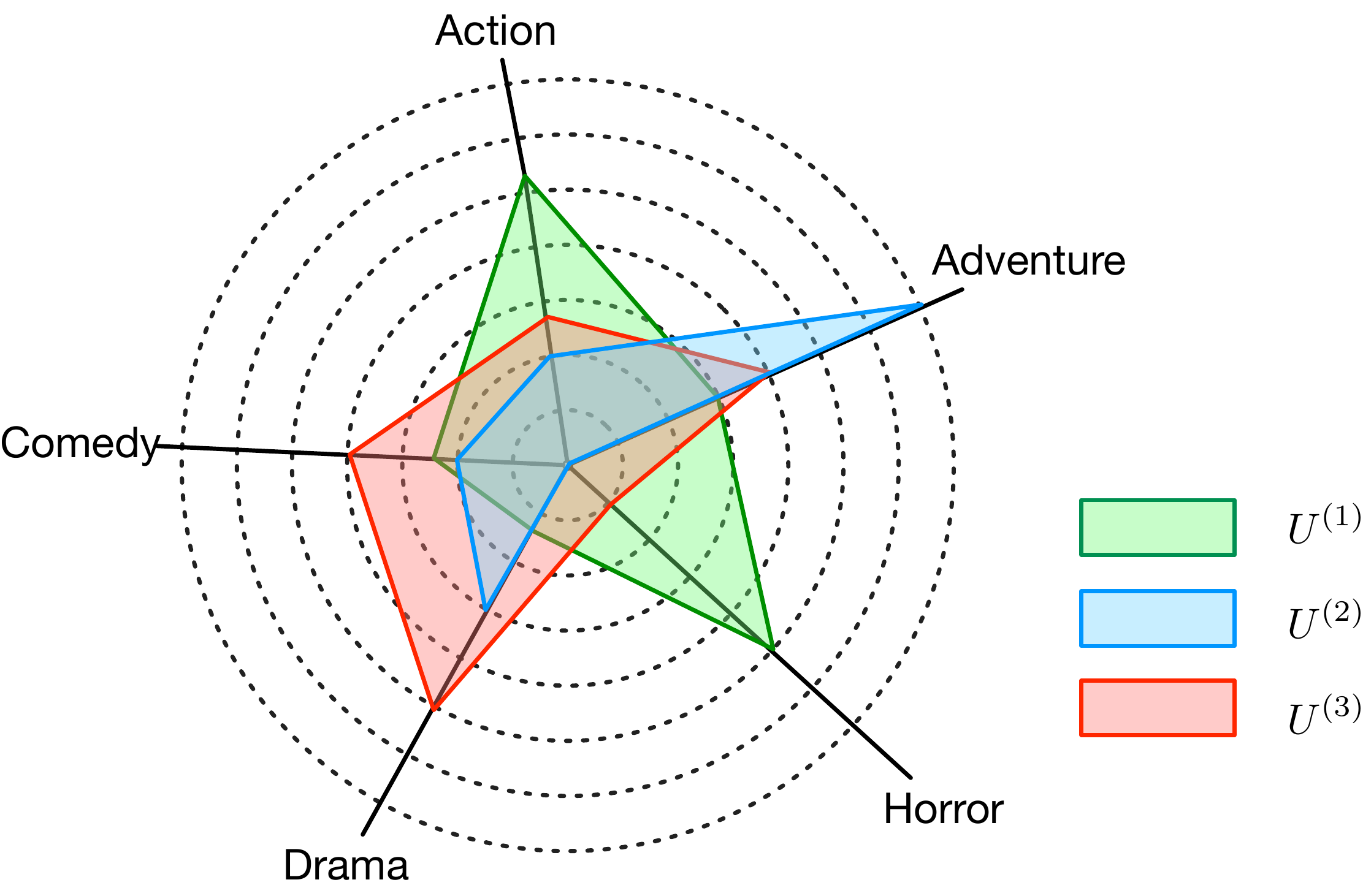}
    \caption{Multiple user preference facets (5) that are used to demonstrate user preference in a movie recommendation task, where each vertex represents a preference degree of the corresponding facet. The figure shows an example of three users.}
    \label{fig:polar_chart}
\end{figure}

\begin{figure*}[t]
  \centering
  \includegraphics[width=0.7\textwidth]{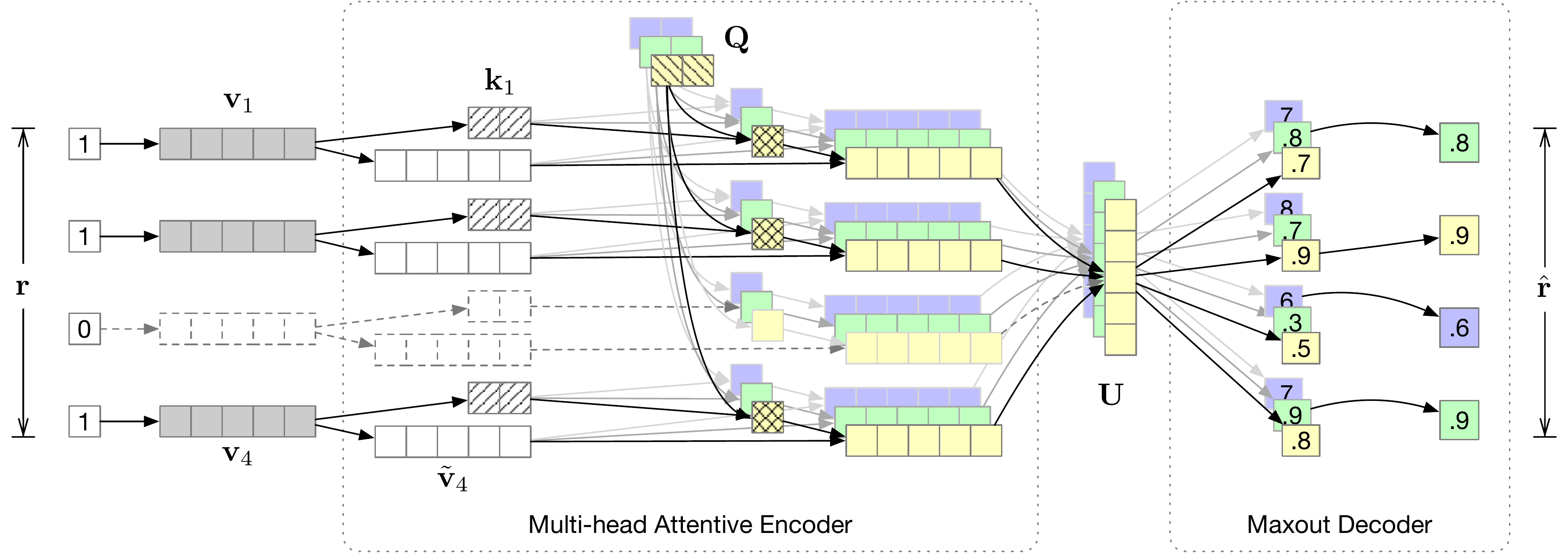}
  \caption{Attentive Multi-modal AutoRec architecture. Colors are used to distinguish preference modes. Gray boxes show fixed embeddings that are not in training scope. Dashed arrow and boxes show the track of dropped items. In this particular example architecture, we have four items in the recommendation scope. The latent embedding dimension is five. Key and query dimensions are both two. Each user has a maximum of three potential preferences. }
  \label{fig:model_structure}
\end{figure*}

\section{Attentive Multi-modal AutoRec}
In this section, we present an alternative attention-based collaborative filtering architecture -- Attentive Multi-modal AutoRec. The proposed model aims to 1) reduce the computation and storage requirements for collaborative filtering, 2) improve the interpretability, and 3) track user preferences from multiple perspectives for each user to improve recommendation performance.

User preferences are often a combination of multiple hidden characteristic facets in recommendation tasks. 
Instead of representing the user preference as a single vector, it is probably more intuitive to represent it as a group of facets, as demonstrated in Figure \ref{fig:polar_chart}. Formally, we want to create an encoding function $f_\vartheta$ such that 
\begin{equation}
    U^{(i)} = f_\vartheta(\mathbf{r}^{(i)}) \quad \textit{and} \quad U^{(i)} = [\mathbf{u}_{1}^{(i)}, \mathbf{u}_{2}^{(i)}\cdots \mathbf{u}_{d}^{(i)}],
\label{eq:multi_mode_encoder}
\end{equation}
where each user $i$ has $d$ diverse preference facets that jointly represent the user's preference. Since the user preference is distributed into multiple vectors, the length of each vector could be substantially reduced.

With multiple user representation vectors, one can make recommendations to the user by jointly considering each of the characteristic facets~(or preference modes) the user has. Formally, there exists a decoder function $f_\theta$ such that:
\begin{equation}
    \hat{r}^{(i)} = f_\theta(U^{(i)}).
\end{equation}

However, in order to achieve our purpose in estimating multiple user preference facets, we still face a few challenges:
\begin{itemize}
    \item How to estimate multiple preference facets when facet side information is not available in the collaborative filtering setting?
    \item How to control the model complexity to train on the real-world dataset with very sparse observations?
\end{itemize}
We describe our approach in detail in the following sub-sections. To provide an overview of our approach, we also show the proposed model architecture in Figure \ref{fig:model_structure}.

\subsection{Multi-head Attention Encoder}
As demonstrated in Equation \ref{eq:multi_mode_encoder}, all user preference facets $\mathbf{u}_l^{(i)}\in U^{(i)}$ are estimated from the same interaction vector $\mathbf{r}^{(i)}$ of user $i$. To allow the preference facets to capture different information, we need to let each preference facet pay attention to a different part of the interactions such that
\begin{equation}
    \mathbf{u}_l^{(i)} = \sum_j a_{j,l}^{(i)}\tilde{\mathbf{v}}_j + \mathbf{b}_l,
\label{eq:multi_mode_attention}
\end{equation}
where $\tilde{\mathbf{v}}_j$ denotes item information that contributes to estimating user facets, $a_{j,l}^{(i)}$ denotes the attention score from user-item intersection $(i,j)$ to the preference facet $l$, and $\mathbf{b}_l$ denotes bias of user preference facet $l$.

For establishing the attention mechanism, we expect each item $j$ has two components: item key $\mathbf{k}_j$ and item value $\tilde{\mathbf{v}}_j$. And, correspondingly, each preference facet $l$ has a query $\mathbf{q}_l$ that is shared for all users. We will describe how we can obtain these values later.
Based on the Scaled Dot-Product Attention described in Section \ref{sec:attention_mechanism}, we can calculate attention $\mathbf{a}^{(i)}_l$ mentioned in Equation \ref{eq:multi_mode_attention} through
\begin{equation}
    \mathbf{a}^{(i)}_l = \frac{1}{Z(\cdot)}\left [r^{(i)}\odot\exp(\frac{q_lK}{\sqrt{\kappa}})\right ],
\end{equation}
where partition function $Z(\cdot)$ sums overall values in the numerator. 

It is important to mention that the attention mechanism is slightly different from its original form. Concretely, we introduce the historical interaction indicator $r^{(i)}$ as a mask in the above function, so that the attention is limited to the observed interactions ($r_j^{(i)} = 1$) for each user.

\noindent{\bf Queries, Keys and Values}
Now we describe how to gather the query, key and value components for the attention computation.

Given item embeddings for all items $V\in\mathbb{R}^{n\times h}$, the key and value for each item are linear transformations of its corresponding item embedding. Formally, we assume there are two linear functions such that:
\begin{equation}
    \mathbf{k}_j = \mathbf{v}_jW_k \quad \textit{and} \quad \tilde{\mathbf{v}}_j = W_v\mathbf{v}_j,
\end{equation}
where coefficients $W_k\in \mathbb{R}^{h\times \kappa}$ and $W_v\in \mathbb{R}^{h\times h}$.

While the queries could also be correlated with potential user embeddings, we believe the preference facet categories for a recommendation task are sharable to all users. Therefore, we simply set the query $q_l\in R^{\kappa}$ as trainable random variables. 

During the model training, the query variables automatically look for the best facet that could represent user preferences. 
Although it is possible to set orthogonal regularization on the query variables to encourage learning diverse preference facets, we note the overlapped facets do not have a significant negative impact on recommendation performance. Thus, we omit our attempts to regularize the queries.

\noindent{\bf Item Embedding}
Since we only track one item embedding matrix $V$ during the encoding stage, we can presume the item embedding matrix comes from various sources without training. However, as we limit our information accessibility to only users' historical interactions, we should obtain the item embedding matrix from the interaction matrix $R$.

One simple solution is the SVD decomposition such that
\begin{equation}
    R = U\Sigma V^T.
\end{equation}
While one can also adopt state-of-the-art approaches such as Graph Convolutional Neural Networks~(GCN) to optimize the item embeddings, we stick to the simple SVD solution to clarify the contributions of our proposed model.

\subsection{Maxout Decoder}
A conventional Autoencoder based recommender system predicts the score of a future interaction by decoding from a user embedding into an item observation space through
\begin{equation}
\hat{\mathbf{r}}^{(i)} = f_\theta(\mathbf{u}^{(i)}).    
\end{equation}

In a simple linear decoding network setting, a dot product between user embedding $\mathbf{u}^{(i)}$ and item-specific decoding weights $\mathbf{s}_j$ are sufficient for the task such that:
\begin{equation}
    r_{j}^{(i)} = {\mathbf{u}^{(i)}}^T\mathbf{s}_j.
\end{equation}
These straightforward methods, while simple, demonstrate strong prediction performance in practice. 

In the multi-modal user preference estimation setting of this paper, we keep the dot product prediction approach and select the maximum prediction score among all predictions from different user preference modes. Specifically, we have
\begin{equation}
    r_{j}^{(i)} = \max(U^{(i)}\mathbf{s}_j),
\end{equation}
where we only maintain one set of decoding weights for all preference facets estimated. The maxout prediction decoder is beneficial from two perspectives:
\begin{itemize}
    \item Items are recommended for different reasons. For example, item A may be recommended because it fit preference facet 1, whereas item B is recommended due to the matching with user preference facet 2.
    \item The model behaviour is easy to interpret. We can easily trace which of the user preference facet dominated the recommendation.
\end{itemize}

\subsection{Training Objective}
As a standard Autoencoder based collaborative filtering method, the proposed model could support various type of objectives from simple Mean Squared Error~(MSE) to the ranking losses such as Bayesian Pair-wise Ranking~(BPR). We choose Weighted MSE in this paper as it is efficient during training and shows reasonable recommendation performance. Formally, the objective function is
\begin{equation}
    {\arg\max}_{\theta,\vartheta} \sum_{i=1}^m \langle \mathbf{c}^{(i)}, (\mathbf{r}^{(i)} - f_\theta(f_\vartheta(\mathbf{\tilde{r}}^{(i)})))^2 \rangle + \lambda\norm{\theta},
\end{equation}
where $\tilde{r}^{(i)}$ is randomly corrupted interaction history and $\mathbf{c}^{(i)}$ is the loss weighting vector of user $i$. Here, the random corruption means we randomly remove some observed interactions of users during each training epoch to reduce the chance of over-fitting. 

While there are multiple choices to produce the loss weighting vector, we use the one described in WRMF~\cite{hu2008collaborative}:
\begin{equation}
    c_{j}^{(i)} = 1 + \alpha\log (1+r_{j}^{(i)}),
\end{equation}
where $\alpha$ is a hyper-parameters.

Recall our previous discussion about the multi-head attention encoder, the encoder parameter set $\vartheta$ consists of several convolution kernels as follows
\begin{equation}
    \vartheta = \{W_k, W_v, Q, B\}
\label{eq:encoder_parameters}
\end{equation}
that are shared to all users and items, which forms a sufficient mutual regularization. As a contrast, the parameter set of the decoder $\theta$ is bounded by the items. 

\begin{equation}
    \theta = \{S\}
\label{eq:decoder_parameters}
\end{equation}
Therefore, we choose to regularize the decoder parameters only.

\subsection{Model Complexity}
The expressiveness and complexity of a machine learning model are usually bounded by the number of parameters involved in the model. Many machine learning based recommender systems make a trade-off between model complexity and the size of available data. 
However, the proposed model demonstrates surprisingly strong performance with a limited number of parameters.

While the proposed model has many parameter sets, as described in Equation \ref{eq:encoder_parameters} and \ref{eq:decoder_parameters}, the total number of parameters are less than classic latent factor models such as matrix factorization. As a concrete example, for the Movielens-1M dataset in our experiments, the proposed AMA model uses $143,169$ parameters to achieve the optimal recommendation performance while matrix factorization, a model with similar performance as AMA, has $1,914,200$ parameters, which is far more than our proposed model. 
Fewer parameters does not 
sacrifice
the modelling flexibility of the proposed model since it supports multiple hyper-parameters. Table \ref{tb:hyper_param_set} in Appendix~\ref{appendix: hyper}\footnote{\label{fn:appendix} Please find the appendix in our extended version on arXiv.} shows all tune-able hyper-parameters of the AMA model.

\section{Experiments and Evaluation}
In this section, we perform experiments aiming at answering the following research questions:
\begin{itemize}
\item {\bf RQ1} How does AMA perform as compared with state-of-the-art CF methods? 

\item {\bf RQ2} How effective is the proposed multi-modal preference estimation?

\item {\bf RQ3} How does the interpretability of the model benefit from the attention mechanism and multi-modal preference representation?

\end{itemize}

\subsection{Experiment Settings}
\noindent{\bf Datasets.} We evaluate the candidate algorithms on three publicly available rating datasets: Movielens-1M, Amazon Digital Music, and  Amazon Video Games. The details of the datasets, preprocessing method and train/test/validation splitting are presented in Appendix~\ref{appendix:data}\footnotemark[\getrefnumber{fn:appendix}].

\noindent{\bf Evaluation protocols.}
We evaluate the recommendation performance using five metrics: Precision@K, Recall@K, MAP@K, R-Precision, and B-NDCG, where R-Precision is an order insensitive metrics, NDCG is order sensitive, and Precision@K as well as Recall@K are semi-order sensitive due to the K values are given.

\noindent{\bf Baselines.}
To demonstrate the effectiveness, we compare our proposed AMA with nine state-of-the-art scalable models with the ability to handle millions of users and transactions. For a fair comparison, we only compare models with similar running time.

\begin{itemize}
\item {\bf POP}: Most popular items -- not user personalized but an intuitive baseline to test the claims of this paper.
\item {\bf AutoRec}~\cite{sedhain2015autorec}: A neural Autoencoder based recommendation system with one hidden layer and ReLU activation function. 
\item {\bf ACF}~\cite{chen2017attentive}: Attentive Collaborative Filtering, which is the most well-known attention based recommender system.
\item {\bf BPR}~\cite{rendle2009bpr}: Bayesian Personalized Ranking. One of the first recommendation algorithms that explicitly optimizes pairwise rankings. 
\item {\bf CDAE}~\cite{wu2016collaborative}: Collaborative Denoising Autoencoder, which is specifically optimized for implicit feedback recommendation tasks. 
\item {\bf CML}~\cite{hsieh2017collaborative}: Collaborative Metric Learning. A state-of-the-art metric learning based recommender system. 
\item {\bf PLRec}~\cite{sanner:ijcai16a,sigir19a}: Projected linear recommendation approach.
\item {\bf PureSVD}~\cite{cremonesi2010performance}: A similarity based recommendation method that constructs a similarity matrix through SVD decomposition of implicit matrix $R$. 
\item {\bf VAE-CF}~\cite{liang2018variational}: Variational Autoencoder for Collaborative Filtering. A state-of-the-art deep learning based recommender system. 

\end{itemize}
More hyper-parameter details are
provided in Appendix~\ref{appendix: hyper}\footnotemark[\getrefnumber{fn:appendix}].

\subsection{Performance Comparison (RQ1)}
In this experiment, we compare the proposed model against several collaborative filtering algorithms on  one-class recommendation tasks. Since some of the recommender systems are not intentionally designed form the OC-CF tasks, we extended and enhanced them for the one-class task from various perspectives. We discuss them in more detail in Appendix~\ref{appendix: algo}\footnotemark[\getrefnumber{fn:appendix}].  

Table \ref{table:ml1m}, \ref{table:digital_music},  and \ref{table:video_games} show the Top-N recommendation performance comparison between the proposed model and various baselines on the one benchmark Movielens-1M dataset and two real-world Amazon datasets. From the tables \footnote{We omit explicit 95\% confidence intervals (CIs) in Tables  \ref{table:ml1m}, \ref{table:digital_music},  and \ref{table:video_games} since all CIs are smaller than $0.001$.}, we obtain the following interesting observations:
\begin{itemize}
    \item Compared to the Autoencoder based recommender systems~(AutoRec++ and CDAE), the proposed Attentive Multi-modal AutoRec model not only shows better performance on the benchmark dataset but also demonstrates better robustness in terms of producing consistently strong recommendations on real data that is extremely sparse and noisy. 
    
    \item Compared to recommender systems that leverage the Noise Contrastive Estimation technique~(ACF-BPR and PLRec), the proposed model shows better recommendation performance. This observation provides concrete evidence that the performance improvement of the proposed model does not merely rely on the quality of prefixed item embeddings.
    
    \item Comparing to the Attention based recommender system~(ACF-BPR and AMA), we note that the proposed model has a consistent advantage on Precision@k metric in general. However, in one of the three datasets, we observe that ACF does better than the proposed model on Recall@K.
    
    \item For the two Amazon datasets, we note the algorithms equipped with BPR objective show significantly better performance than the other recommender systems. This observation indicates the practical advantage of the methods that directly optimize ranking. However, to show the contribution of the proposed novel architecture, we do not let the proposed model take advantage of BPR loss.
    
\end{itemize}

\begin{table*}[t!]
\caption{Results of Movielens-1M dataset. Hyper-parameters are chosen from the validation set.}
\resizebox{\textwidth}{!}{%
\begin{tabular}{ccccccccccccc}
\toprule
Model&R-Precision&NDCG&MAP@5&MAP@10&MAP@20&Precision@5&Precision@10&Precision@20&Recall@5&Recall@10&Recall@20\\
\midrule
POP&7.36\%&12.25\%&11.68\%&10.99\%&10.31\%&10.85\%&9.99\%&9.34\%&2.37\%&4.34\%&8.41\%&\\
AutoRec++&9.45\%&16.74\%&12.54\%&12.02\%&11.31\%&11.94\%&11.15\%&10.29\%&3.77\%&6.90\%&12.15\%\\
ACF-BPR&8.40\%&14.56\%&11.54\%&11.10\%&10.47\%&11.17\%&10.37\%&9.46\%&3.14\%&5.55\%&9.98\%\\
MF-BPR&9.33\%&16.75\%&11.92\%&11.53\%&11.02\%&11.41\%&10.94\%&10.16\%&3.75\%&6.92\%&12.28\%\\
CDAE&9.41\%&15.88\%&12.97\%&12.34\%&11.55\%&12.26\%&11.43\%&10.35\%&3.50\%&6.18\%&10.79\%\\
CML&10.00\%&17.67\%&{\bf13.56\%}&12.99\%&{\bf12.29\%}&12.97\%&12.21\%&{\bf11.14\%}&4.00\%&7.18\%&12.65\%\\
PLRec&9.91\%&17.90\%&13.10\%&12.50\%&11.81\%&12.48\%&11.70\%&10.71\%&{\bf4.24\%}&7.75\%&13.58\%\\
PureSVD&9.20\%&16.44\%&12.12\%&11.61\%&10.98\%&11.60\%&10.79\%&9.98\%&3.83\%&6.93\%&12.30\%\\
VAE-CF&8.92\%&16.34\%&10.66\%&10.45\%&10.06\%&10.54\%&10.08\%&9.42\%&3.76\%&6.88\%&12.07\%\\
\midrule
AMA&{\bf10.14\%}&{\bf17.91}\%&13.44\%&{\bf13.00\%}&12.27\%&{\bf13.02\%}&{\bf12.25\%}&11.12\%&4.18\%&7.62\%&13.25\%\\
\bottomrule
\end{tabular}}
\label{table:ml1m}
\end{table*}

\begin{table*}[t!]
\caption{Results of Amazon Digital Music dataset. Hyper-parameters are chosen from the validation set.}
\resizebox{\textwidth}{!}{%
\begin{tabular}{ccccccccccccc}

\toprule
Model&R-Precision&NDCG&MAP@5&MAP@10&MAP@20&Precision@5&Precision@10&Precision@20&Recall@5&Recall@10&Recall@20\\
\midrule
POP&0.29\%&1.10\%&0.29\%&0.27\%&0.23\%&0.28\%&0.22\%&0.17\%&0.64\%&0.92\%&1.53\%\\
AutoRec++&0.34\%&1.20\%&0.36\%&0.31\%&0.26\%&0.29\%&0.24\%&0.19\%&0.64\%&1.09\%&1.68\%\\
ACF-BPR&2.99\%&{\bf 7.15}\%&2.39\%&1.91\%&1.48\%&1.74\%&1.27\%&0.92\%&{\bf 5.25}\%&{\bf 7.26}\%&{\bf 10.14}\%\\
MF-BPR&3.04\%&6.94\%&2.45\%&1.92\%&1.47\%&1.71\%&1.23\%&0.89\%&5.23\%&7.11\%&9.88\%\\
CDAE&0.32\%&1.17\%&0.33\%&0.30\%&0.25\%&0.32\%&0.24\%&0.18\%&0.70\%&1.09\%&1.65\%\\
CML&0.29\%&1.50\%&0.31\%&0.28\%&0.25\%&0.26\%&0.24\%&0.23\%&0.60\%&1.05\%&2.16\%\\
PLRec&1.62\%&4.62\%&1.52\%&1.25\%&1.00\%&1.19\%&0.89\%&0.68\%&3.05\%&4.46\%&6.79\%\\
PureSVD&1.61\%&4.06\%&1.47\%&1.17\%&0.90\%&1.07\%&0.78\%&0.56\%&2.84\%&4.04\%&5.71\%\\
VAE-CF&2.94\%&6.68\%&2.32\%&1.83\%&1.41\%&1.67\%&1.21\%&0.86\%&4.99\%&6.94\%&9.43\%\\
\midrule
AMA&{\bf3.08\%}&6.90\%&{\bf2.59\%}&{\bf2.05\%}&{\bf1.57\%}&{\bf1.85\%}&{\bf1.33\%}&{\bf0.94\%}&5.06\%&7.04\%&9.60\%\\
\bottomrule
\end{tabular}}
\label{table:digital_music}
\end{table*}

\begin{table*}[t!]
\caption{Results of Amazon Video Games dataset. Hyper-parameters are chosen from the validation set.}
\resizebox{\textwidth}{!}{%
\begin{tabular}{ccccccccccccc}
\toprule
Model&R-Precision&NDCG&MAP@5&MAP@10&MAP@20&Precision@5&Precision@10&Precision@20&Recall@5&Recall@10&Recall@20\\
\midrule
POP&0.24\%&1.12\%&0.21\%&0.19\%&0.19\%&0.17\%&0.19\%&0.17\%&0.46\%&0.98\%&1.74\%\\
AutoRec++&0.31\%&1.81\%&0.37\%&0.36\%&0.32\%&0.39\%&0.32\%&0.25\%&1.04\%&1.69\%&2.73\%\\
ACF-BPR&1.19\%&4.91\%&1.06\%&0.95\%&0.83\%&0.93\%&0.80\%&0.65\%&2.90\%&4.83\%&7.59\%\\
MF-BPR&1.48\%&5.41\%&1.27\%&1.11\%&0.94\%&1.07\%&0.88\%&0.71\%&3.23\%&5.13\%&8.19\%\\
CDAE&0.35\%&1.62\%&0.36\%&0.32\%&0.28\%&0.33\%&0.25\%&0.23\%&0.85\%&1.32\%&2.48\%\\
CML&1.18\%&3.79\%&0.98\%&0.81\%&0.67\%&0.74\%&0.59\%&0.49\%&2.08\%&3.26\%&5.33\%\\
PLRec&1.94\%&5.69\%&1.69\%&1.41\%&1.14\%&1.34\%&1.04\%&0.77\%&3.95\%&5.96\%&8.61\%\\
PureSVD&1.32\%&4.08\%&1.16\%&0.98\%&0.79\%&0.93\%&0.72\%&0.54\%&2.81\%&4.23\%&6.09\%\\
VAE-CF&1.82\%&5.62\%&1.49\%&1.25\%&1.02\%&1.19\%&0.92\%&0.71\%&3.79\%&5.71\%&8.51\%\\
\midrule
AMA&{\bf2.16\%}&{\bf6.62\%}&{\bf1.83\%}&{\bf1.54\%}&{\bf1.25\%}&{\bf1.46\%}&{\bf1.13\%}&{\bf0.88\%}&{\bf4.48\%}&{\bf6.76\%}&{\bf10.18\%}\\
\bottomrule
\end{tabular}}
\label{table:video_games}
\end{table*}


\subsection{Effect of Multi-modal Estimation (RQ2)}
To better evaluate the effectiveness of the multi-modal estimation in our proposed model, we will try to answer the following questions: (a) How many users benefit from using more than one user preference?  (b) How does the number of user preferences estimated affect the Precision and Recall of the model? (c) What are the characteristics of the top-10 items that each preference mode pays the greatest attention to?

\begin{figure}[t]
  \centering
    \begin{subfigure}{0.45\linewidth}
    \includegraphics[width=\linewidth]{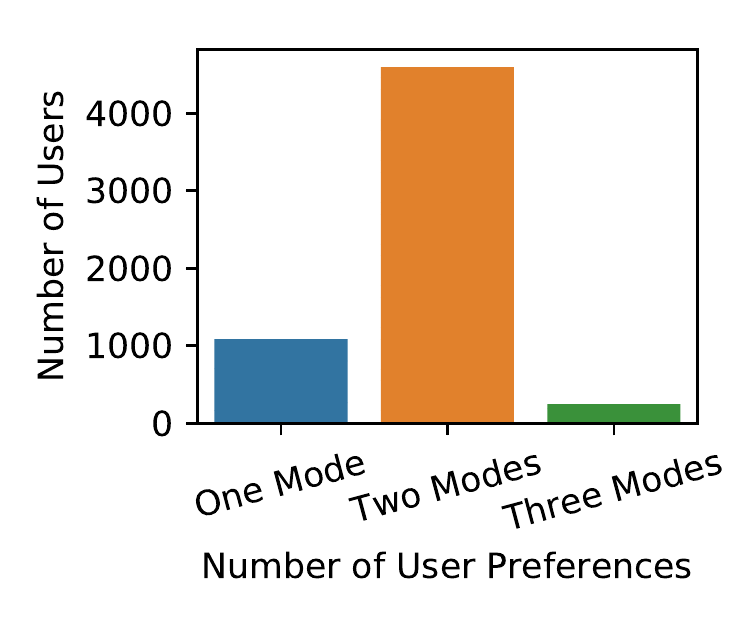}
    \caption{MovieLens}
    \end{subfigure}
    \begin{subfigure}{0.45\linewidth}
    \includegraphics[width=\linewidth]{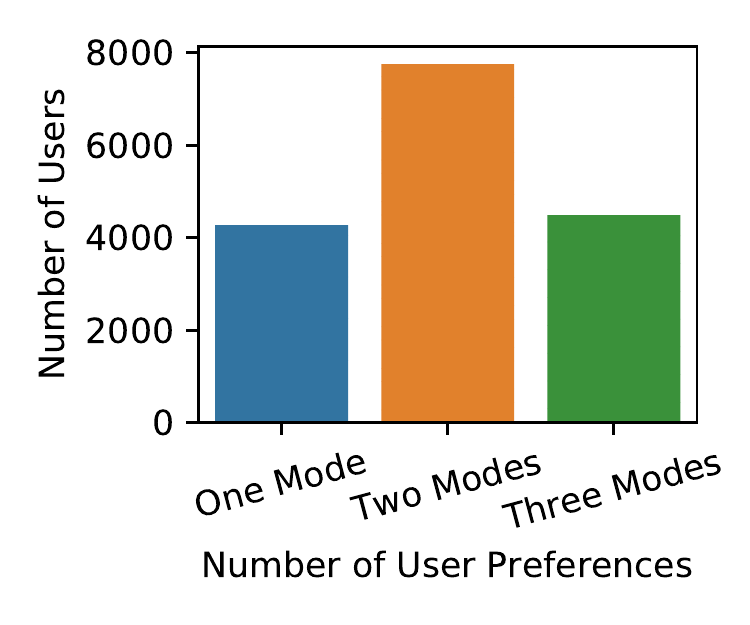}
    \caption{Amazon Digital Music}
    \end{subfigure}
    \caption{Statistics of the number of user preference modes being used for producing the Top-10 recommendations for each user. }
  \label{fig:modes_analysis}
\end{figure}

\begin{figure}[t]
    \centering
    \includegraphics[width=0.75\linewidth]{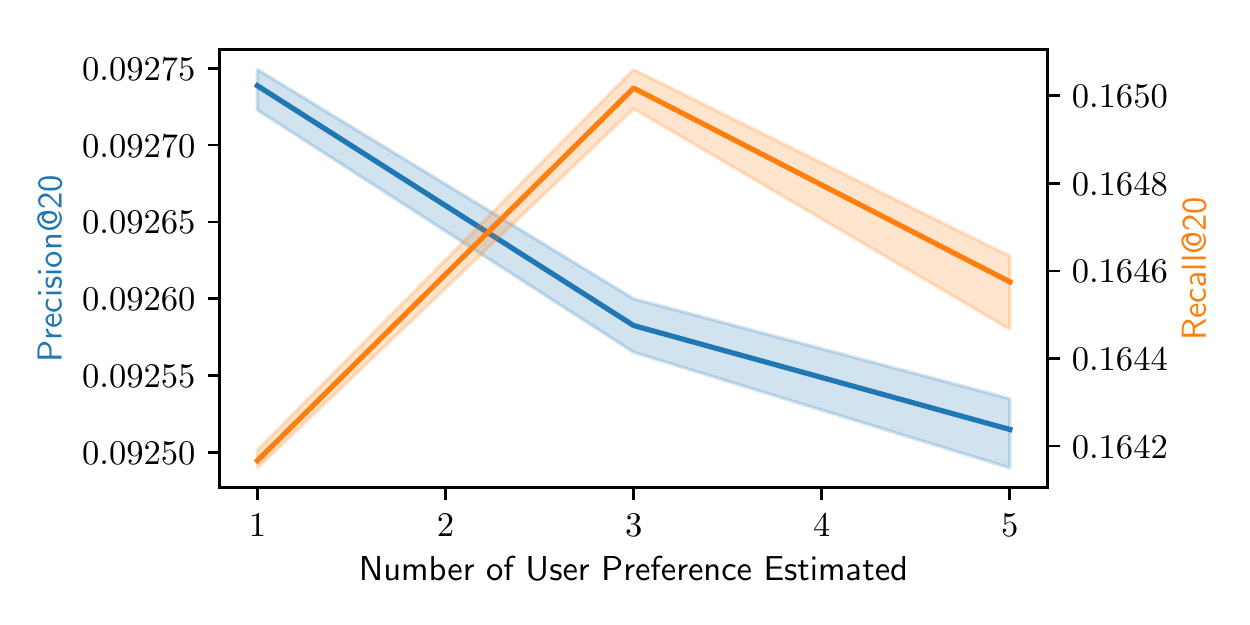}
    \caption{Recommendation performance varies based on the number of user preference modes or facets that are explicitly estimated.}
    \label{fig:modes_analysis_2}
\end{figure}

\begin{enumerate}
    \item As Figure \ref{fig:modes_analysis} depicts, in a setting where three user preference modes are applied, more than 82\% of users in MovieLens-1M and 74\% of users in Amazon Digital Music take advantage of more than one user preference mode during recommendation. The observation reflects our hypothesis that most of the users have more than one distinct preference mode and could benefit from explicit modelling of multiple preference modes.
    \item In order to understand the influence of modeling multiple preference modes on the overall recommendation performance, we compare the AMA models using one, three, and five prefixed number of modes. As shown in Figure \ref{fig:modes_analysis_2}, the AMA model achieves the best Precision@K score when it is initialized with a single mode. In contrast, it achieves the best Recall@K score when it leverages three modes. From this perspective, the AMA model provides a way to control the precision-recall trade-off depends on different scenario requirements.
    \item If we take the Movielens-1M dataset as an example, it is easy to find that each preference mode captures one or more movie genres. Although the preferences are not explicitly bounded with any side information, they are automatically captured by the attention mechanism. The details of the analysis for all datasets can be found in Appendix~\ref{appendix: case}\footnotemark[\getrefnumber{fn:appendix}]. 
    
\end{enumerate}

\begin{figure*}[t]
  \centering
  \includegraphics[width=0.43\linewidth]{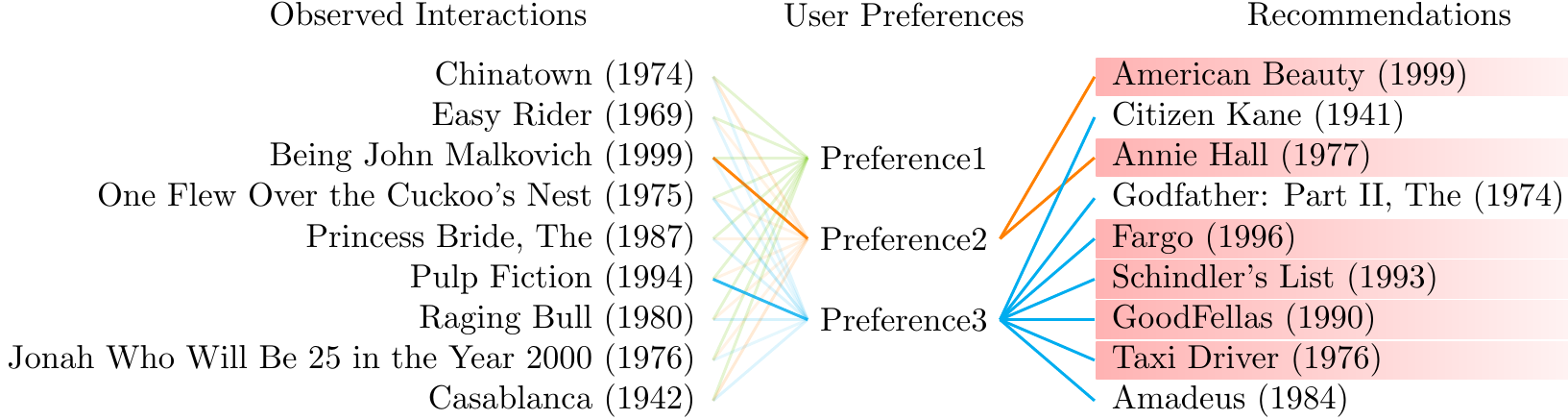}
  \includegraphics[width=0.56\linewidth]{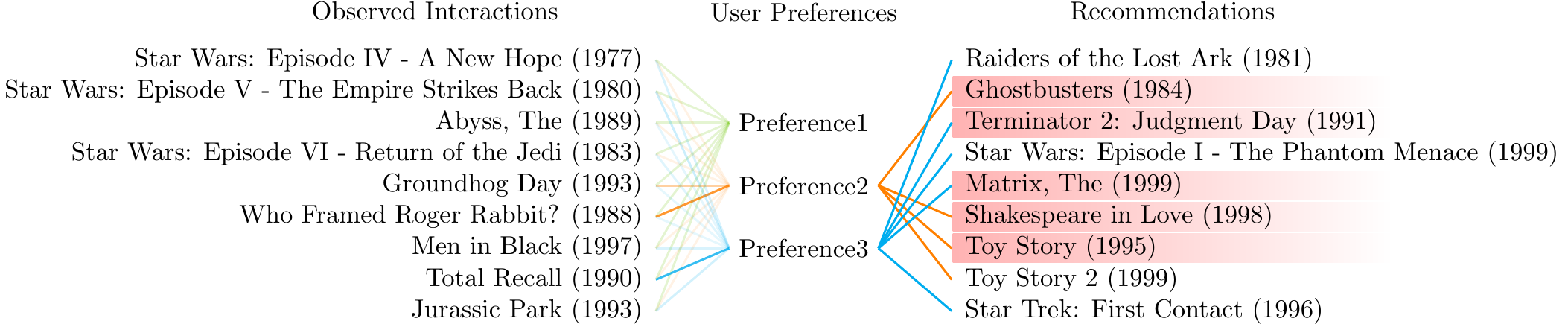}
  \caption{Attention case study for users with multiple active preference modes. Red entries in the recommendation lists represent the recommendations hit ground truth interactions in the test set.}
  \label{fig:demo1}
\end{figure*}

\begin{figure*}[t]
  \centering
  \includegraphics[width=0.51\linewidth]{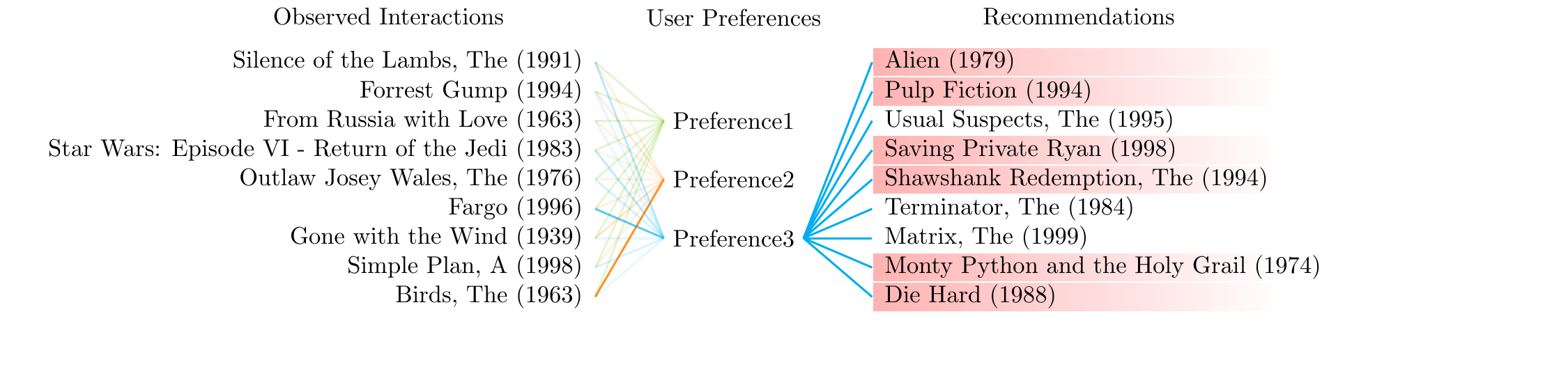}
  \includegraphics[width=0.46\linewidth]{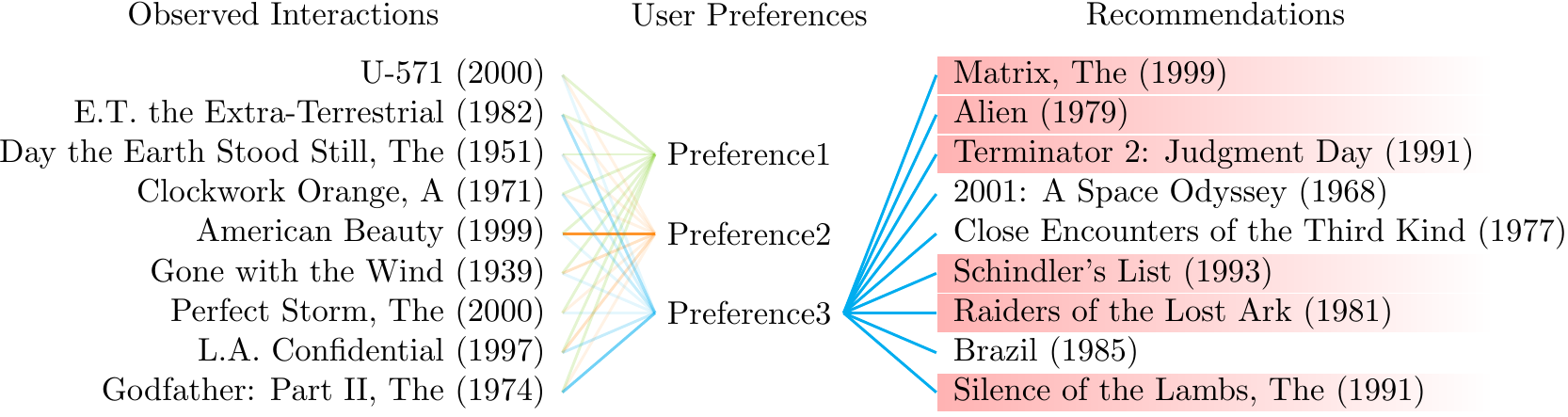}
  \caption{Attention case study for users with single active preference mode. Red entries in the recommendation lists represent the recommendations hit ground truth interactions in the test set.}
  \label{fig:demo2}
\end{figure*}

\subsection{Attentive Interpretation Analysis (RQ3)}
A key advantage of AMA is the improved interpretability since we can visualize how AMA generates recommendation. With an attention mechanism, we have the weighted importance of each observed interaction for different user embedding vectors and with the Maxout Decoder, we know which user embedding vector a recommendation comes from. This not only helps us to understand how the model is learned but also provides a reasonable explanation of the recommendation. 

We present qualitative examples for four users from Movielens-1M to demonstrate the interpretability of our proposed model.  In each example, there are three columns: apart from observed interactions and recommendations, we have user preferences in the middle, and each user preference represents one user embedding vector. The lines between the movies and each user preference represent the weighted importance; the more solid the line, the more important the movie is to this preference. The line between a user preference and the recommendation indicates that the movie is recommended based on this user preference. The red entries in the recommendation lists represent the recommendations that ``hit'' the ground truth interactions in the test set. 

According to the experimental results, we find that different user preferences recommend very diverse movies. For the user in Figure \ref{fig:demo1}(a), the criminal movie Pulp Fiction contributes the most to the third user preference, which accordingly recommends a variety of related criminal movies such as GoodFellas and Taxi Driver. At the same time, the second preference learned to pay more attention to comedy-drama movies such as Being John Malkovich, and this preference recommends Annie Hall, another famous comedy-drama movie. Both the examples in Figure \ref{fig:demo1}(a) and \ref{fig:demo1}(b) show how we can interpret a recommendation and observe the effectiveness of attention for multi-modal preference estimation.

It is worth mentioning that compared with the traditional AutoRec, AMA is more flexible and can be treated as a general form of AutoRec since AMA can decide to use a different number of preferences according to different users. As shown in Figure \ref{fig:demo1}, for users with a broader range of interest, the model tends to use two preferences for recommendation, while in Figure \ref{fig:demo2}, the model chooses to use just one preference for users with less diverse taste. 


\section{Conclusion}
In this paper, we proposed a novel and efficient model called Attentive Multi-modal AutoRec (AMA) to explicitly track multiple facets of user preferences by introducing multi-modal estimation into AutoRec. To do this, AMA leverages an attention mechanism to assign different attention weights to each item interaction when it is used to estimate different preference facets of the user. 
AMA not only captures multifaceted user preferences in a latent space but also improves the interpretability of the model since we know which latent preference facet a recommendation comes from and which items contribute to the corresponding user preference facet. Through extensive experiments on three real-world datasets, we demonstrated the effectiveness of the multi-modal estimation in our model and showed that AMA is competitive with many state-of-the-art models in the OC-CF setting. 

\bibliographystyle{IEEEtran}
\bibliography{IEEEabrv,IEEEexample}

\newpage
\appendices
\counterwithin{table}{section}

\section{Hyper-parameter detail}\label{appendix: hyper}
In Table~\ref{tb:hyper_param_set}, we show all the tune-able parameters of our AMA model.
We tune the hyper-parameters for all the candidate algorithms by evaluating on the validation datasets through the greedy search.
The best hyper-parameter settings found for each algorithm and domain are listed in Table~\ref{tb:hyper-parameter}.
\begin{table}[h!]
  \caption{Hyper-parameter set for AMA model}
  \resizebox{0.5\textwidth}{!}{%
  \begin{tabular}{c|lc}
  \toprule
    Symbol &\multicolumn{1}{c}{Function}&\\
  \midrule
  $h$& Size of latent representation vector \\
  $d$& Number of user preference facets \\
  $\kappa$ & Size of key and query vector \\
  $\alpha$ & Loss weighting parameter\\
  $\rho$ & Random input corruption rate (see \cite{vincent2008extracting})\\
  $NCE$ & Noise Contrastive Estimation based item embedding \\
  & initialization (see \cite{sigir19a}) \\
    \hline
  \end{tabular}
  }
 \label{tb:hyper_param_set}
\end{table}
\begin{table}[h!]
 \caption{Best hyper-parameter setting for each algorithm.}
  \resizebox{\linewidth}{!}{%
  \begin{threeparttable}
  \begin{tabular}{c|ccccccccc}
  \toprule
Domain&Algorithm&$h$&$\alpha$&$\lambda$&$\upvarepsilon$&$\upgamma$&$\rho$&$d$& NCE?\\
\midrule
&POP&-&-&-&-&-&-&-&-\\
&AutoRec&200&-&1E-05&300&-&-&-&-\\
&ACF-BPR&100&-&0.0001&100&10&-&-&\checkmark\\
&MF-BPR&200&-&1E-05&30&-&-&-&-\\
&CDAE&200&-&1E-05&300&-&0.5&-&-\\
MovieLens-1M&CML&200&-&0.1&30&-&-&-&-\\
&PLRec&100&-&10000&10&10&-&-&\checkmark\\
&PureSVD&50&-&1&10&10&-&-&-\\
&VAE-CF&200&-&1E-05&300&-&0.4&-&-\\
&AMA&40&1&1e-05&300&10&0.3&3&\checkmark\\
\midrule
&POP&-&-&-&-&-&-&-&-\\
&AutoRec&200&-&0.0001&300&-&-&-&-\\
&ACF-BPR&200&-&1E-05&100&10&-&-&\checkmark\\
&MF-BPR&200&-&0.0001&30&-&-&-&-\\
Amazon&CDAE&200&-&0.0001&300&-&0.3&-&-\\
Digital Music&CML&200&-&0.001&30&-&-&-&-\\
&PLRec&200&-&10000&-&10&-&-&\checkmark\\
&PureSVD&200&-&-&-&10&-&-&-\\
&VAE-CF&200&-&1E-05&300&-&0.2&-&-\\
&AMA&200&10&0.0001&300&10&0.4&5&\checkmark\\
\midrule
&POP&-&-&-&-&-&-&-&-\\
&AutoRec&100&-&0.0001&300&-&-&-&-\\
&ACF-BPR&200&-&0.0001&100&10&-&-&\checkmark\\
&MF-BPR&200&-&0.0001&30&-&-&-&-\\
Amazon&CDAE&100&-&1E-05&300&-&0.4&-&-\\
Video Games&CML&200&-&0.01&30&-&-&-&-\\
&PLRec&200&-&10000&-&10&-&-&\checkmark\\
&PureSVD&100&-&1&-&10&-&-&-\\
&VAE-CF&200&-&1E-05&300&-&0.2&-&-\\
&AMA&100&10&0.001&300&10&0.4&1&\checkmark\\

    \hline
  \end{tabular}
  \begin{tablenotes}[para,flushleft]
    * In this table, $\lambda$: regularization $\upvarepsilon$: epoch, $\upgamma$: optimization iteration\\
    * For PureSVD, PLRec and AMA, optimization iteration means number of randomized SVD iterations. For WRMF, iteration shows number of alternative close-form optimizations.
\end{tablenotes}
  \end{threeparttable}
  }%
\label{tb:hyper-parameter}
\end{table}

\section{Dataset}\label{appendix:data}
For each dataset, we binarize the rating dataset with a rating threshold, $\vartheta$, defined as the upper half of the range of the ratings. We do this so that the observed interactions correspond to positive feedback. To be specific,  the threshold is $\vartheta > 3$ for all datasets. Table \ref{tb:dataset} summarizes the properties of the binarized matrices. 


We split the data into train, validation and test sets based on timestamps given by the dataset to provide a recommendation evaluation setting closer to production use~\cite{shani2011evaluating}. For each user, we use the first 50\% of data as the train set, 20\% data as validation set and 30\% data as the test set.
\begin{table}
\caption{Datasets statistics.}
  \resizebox{\linewidth}{!}{%
  \begin{tabular}{cccccc}
  \toprule
    Dataset & $m$ & $n$ & $|r_{i,j}>\vartheta|$& Sparsity&\\
   \midrule
    MovieLens-1M & 6,038&3,533&575,281&$2.7\times10^{-2}$\\
    Amazon Digital Music & 16,502& 11,795 & 136,858&$7.0\times10^{-4}$\\
    Amazon Video Games & 54,721& 17,365 & 378,949&$4.0\times10^{-4}$\\
    \hline
  \end{tabular}
  }
 \label{tb:dataset}
\end{table}

\section{What are the top-10 items that each preference mode pays the greatest attention to?}\label{appendix: case}
Table \ref{table:ml_preference_analysis}, \ref{table:music_preference_analysis}, and \ref{table:video_preference_analysis} show the top-10 items that receives the highest attention for each of the three preference modes. 
For the Movielens-1M dataset in Table \ref{table:ml_preference_analysis}, it is easy to note that the preference mode 1 captures the sci-fi movies, preference mode 2 captures comedy and drama, and preference mode 3 captures horror movies. While the preferences are not explicitly bounded with any side information, they are automatically captured by the attention mechanism. 

However, we also note that not all of the preference modes are interpretable. The three preferences captured in table \ref{table:music_preference_analysis}, while having a noticeable difference in item popularity, are hard to summarize using natural language. 

In Table \ref{table:video_preference_analysis}, we see that two out of three preference modes capture various outliers that have lower item popularity than top ranked items in another mode. This observation is consistent with our hyper-parameter tuning results shown in Table \ref{tb:hyper-parameter}, where the AMA performs the best on the Amazon Video Game dataset when it selects only one mode. These results show that not all datasets benefit from multi-modal user preference estimation and for datasets where uni-modal estimation is sufficient, AMA is able to adapt based on the characteristics of the datasets.
\begin{table*}[ht]
\caption{Top-10 movies that receive the highest attentions for the three preference modes respectively (Movielens-1M)}
\vspace{-1mm}
\resizebox{\textwidth}{!}{%
\begin{tabular}{l|c|c|l|c|c|l|c|c}
\toprule
\multicolumn{3}{c|}{Preference1}&\multicolumn{3}{c|}{Preference2}&\multicolumn{3}{c}{Preference3}\\
\midrule
\multicolumn{1}{c|}{Movie Name} & \multicolumn{1}{c|}{Rank} & \multicolumn{1}{c|}{Count} &\multicolumn{1}{c|}{Movie Name}& \multicolumn{1}{c|}{Rank} & \multicolumn{1}{c|}{Count} &\multicolumn{1}{c|}{Movie Name}& \multicolumn{1}{c|}{Rank} & \multicolumn{1}{c}{Count}\\
\midrule
Terminator, The (1984)&1&	1015&	Clerks (1994)&1&	628&	Edward Scissorhands (1990)&1&	476\\
Alien (1979)&2&	1094&	Being John Malkovich (1999)&2&	1270&	Professional, The (1994)&2&	366\\
Matrix, The (1999)&3&	1490&	American Pie (1999)&3&	527&	Crying Game, The (1992)&3&	403\\
Aliens (1986)&4&	887&	South Park: Bigger, Longer and Uncut (1999)&4&	493&	Sense and Sensibility (1995)&4&	359\\
Terminator 2: Judgment Day (1991)&5&	1419&	Election (1999)&5&	740&	Like Water for Chocolate (1992)&5&	297\\
Star Wars: Episode IV - A New Hope (1977)&6&	1894&	Austin Powers: The Spy Who Shagged Me (1999)&6&	421&	Interview with the Vampire (1994)&6&	171\\
Star Wars: Episode V - The Empire Strikes Back (1980)&7&	1839&	Wrong Trousers, The (1993)&7&	585&	Nightmare Before Christmas, The (1993)&7&	314\\
Die Hard (1988)&8&	722&	Close Shave, A (1995)&8&	412&	Rocky Horror Picture Show, The (1975)&8&	302\\
Fugitive, The (1993)&9&	955&	Weird Science (1985)&9&	80&	Much Ado About Nothing (1993)&9&	241\\
Total Recall (1990)&10&	770&	Office Space (1999)&10&	275&	Room with a View, A (1986)&10&	189\\

\bottomrule
\end{tabular}}
\label{table:ml_preference_analysis}
\end{table*}

\begin{table*}[ht]
\caption{Top-10 albums that that receive the highest attentions for the three preference modes respectively. (Digital Music)}
\resizebox{\textwidth}{!}{%
\begin{tabular}{l|c|c|l|c|c|l|c|c}
\toprule
\multicolumn{3}{c|}{Preference1}&\multicolumn{3}{c|}{Preference2}&\multicolumn{3}{c}{Preference3}\\
\midrule
\multicolumn{1}{c|}{Album Name} & \multicolumn{1}{c|}{Rank} & \multicolumn{1}{c|}{Count} &\multicolumn{1}{c|}{Album Name}& \multicolumn{1}{c|}{Rank} & \multicolumn{1}{c|}{Count}&\multicolumn{1}{c|}{Album Name}& \multicolumn{1}{c|}{Rank} & \multicolumn{1}{c}{Count}\\
\midrule
Happy by Various artists&1&376&Danza Kuduro by Don Omar&1&31&Blurred Lines &1&196\\
Radioactive by Imagine Dragons&2&168&American VI: Ain't No Grave by Johnny Cash&2&22&Royals&2&100\\
A Thousand Years by Christina Perri&3&129&Uncaged by Zac Brown Band&3&27&Roar by Katy Perry&3&151\\
Somebody That I Used To Know by Gotye&4&129&Court Yard Hounds&4&15&When I Was Your Man by Bruno Mars &4&80\\
Over The Rainbow/What A Wonderful World by Israel&5&94&Sultans of Swing by Dire Straits&5&22&Let It Go by Idina Menzel &5&115\\
Let Her Go by Passenger&6&109&Give Your Heart a Break by Demi Lovato&6&31&Skyfall by Adele&6&85\\
Break Every Chain by Tasha Cobbs Leonard&7&88&Charleston, SC 1966 by Darius Rucker &7&19&Moves Like Jagger by Maroon 5&7&119\\
Oh Honey by Delegation&8&12&Amazing by Ricky Dillard \& New G&8&19&Sail by AWOLNATION&8&93\\
I Loved Her First by The Heartland&9&31&You \& I by Avant&9&25&Honest Face by Liam Finn + Eliza Jane&9&89\\
War by Charles Jenkins \& Fellowship Chicago&10&35&The One by Eric Benét&10&21&Thinking out Loud by Ed Sheeran&10&108\\

\bottomrule
\end{tabular}}
\label{table:music_preference_analysis}
\end{table*}

\begin{table*}[ht]
\caption{Top-10 games that that receive the highest attentions for the three preference modes respectively. (Video Games)}
\resizebox{\textwidth}{!}{%
\begin{tabular}{l|c|c|l|c|c|l|c|c}
\toprule
\multicolumn{3}{c|}{Preference1}&\multicolumn{3}{c|}{Preference2}&\multicolumn{3}{c}{Preference3}\\
\midrule
\multicolumn{1}{c|}{Game Name} & \multicolumn{1}{c|}{Rank} & \multicolumn{1}{c|}{Count} &\multicolumn{1}{c|}{Game Name}& \multicolumn{1}{c|}{Rank} & \multicolumn{1}{c|}{Count}&\multicolumn{1}{c|}{Game Name}& \multicolumn{1}{c|}{Rank} & \multicolumn{1}{c}{Count}\\
\midrule
Final Fantasy VII: Sony PlayStation&1&205&The Bourne Conspiracy - PlayStation 3&1&2&Turtle Beach Ear Force PX22 Gaming Headset&1&50\\
Grand Theft Auto: Vice City: Sony PlayStation 2&2&182&Guitar Hero World Tour Drum - Nintendo Wii &2&3&Turtle Beach Ear Force PX22 Gaming Headset&2&103\\
The Legend of Zelda: The Wind Waker: Nintendo&3&140&Datel Tool Battery PSP Slim&3&1&The Wolf Among Us - Xbox 360 &3&2\\
Mass Effect - Xbox 360&4&200&Teenage Zombies: Nintendo DS&4&1&Snoopy's Grand Adventure - Xbox 360&4&1\\
Call of Duty: Modern Warfare 3 - Xbox 360 &5&181&Battle of the Bands - Nintendo Wii&5&1&Back To The Future 30th Anniversary Xbox One&5&3\\
Resistance: Fall Of Man - PlayStation 3&6&119&Bensussen Deutsch Power Wired Controller for PS3&6&2&Back To The Future 30th Anniversary Xbox 360 &6&3\\
Buy Halo 2 - Xbox with Ubuy Kuwait&7&200&Lara Croft Tomb Raider: Anniversary&7&1&BlazBlue: Chrono Phantasma EXTEND - Xbox One&7&6\\
Far Cry 3 X360&8&196&Master Of Orion 2 - Battle at Antares - PC&8&2&The Amazing Spider-man vs The Kingpin - Sega&8&2\\
Resident Evil 4 - Wii: nintendo wii&9&164&PLAY-IN-CASE CLASSIC for PSP SLIM&9&2&Sunset Overdrive - Xbox One Digital Code Xbox One&9&2\\
New Super Mario Bros&10&251&Gaming Headset For PS4 Xbox One Smart Phone&10&1&Dynamite Cop - Sega&10&3\\
\bottomrule
\end{tabular}}
\label{table:video_preference_analysis}
\end{table*}
\section{Baselines modification}\label{appendix: algo}
We compare our proposed model against several collaborative filtering algorithms on  one-class recommendation tasks. Since some of the recommender systems are not intentionally designed form the OC-CF tasks, we extended and enhanced them for the one-class task from various perspectives. For example, we modified AutoRec in two aspects: 1) we replace Sigmoid activation function with Relu activation, 2) we replace the Mean Square Error with Sigmoid Cross-entropy loss. Both of these modifications show reasonable performance improvement and even outperform some state-of-the-art algorithms on the Movielens-1M benchmark dataset. In addition, we enhanced the ACF-BPR and PLRec algorithms with the recently introduced NCE technique~\cite{sigir19a} and observed remarkable performance improvement over its vanilla form with item embeddings from an SVD decomposition.
\end{document}